\documentclass[preprint,aps,prl,showpacs,preprintnumbers,unsortedaddres,superscriptaddress,amsmath,amssymb,showkeys,floatfix]{revtex4-2}
\usepackage{graphicx}

\usepackage{physics}
\usepackage{xcolor}

\begin{document}

\title{Level rearrangement in $K^- p$ system}

\author{T. Massimino}
\affiliation{%
Department of Physics, California State University, Long Beach, California 90840, USA }%

\author{N.V. Shevchenko}%
 \email{shevchenko@ujf.cas.cz}
\affiliation{%
 Nuclear Physics Institute of the Czech Academy of Sciences, 25068 \v{R}e\v{z}, Czech Republic }%

\author{Z. Papp}
\affiliation{%
Department of Physics, California State University, Long Beach, California 90840, USA }%

\author{J. R\'evai}
\affiliation{%
Budakeszi, H-2092, Hungary}%

\date{\today}

\begin{abstract}
We studied the level shifts in the $K^- p$ system caused by the interplay of strong nuclear and long range Coulomb
potentials. We observed a level rearrangement in the system and found that the $1s$ shift of kaonic hydrogen
is in fact ``attractive''. In addition, we demonstrated that absorption in the strong antikaon-nucleon interaction
does not destroy the level rearrangement. 
\end{abstract}



\maketitle

To our knowledge, the antikaon-nucleon system is the only system where the strong and Coulomb interactions can
separately  lead to the existence of bound or quasi-bound states.  The pure Coulomb attraction between negatively
charged $K^- $ and positively charged $p$ leads  to the existence of an exotic atom, kaonic hydrogen.  In this case,
the strong $\bar{K}N$ interaction plays a secondary role, shifting the lowest levels and giving them width.  On the other
hand, the $\Lambda(1405)$ resonance is usually considered as a quasi-bound state of an antikaon and nucleon which
interacts by the strong force and can decay into a $\pi \Sigma$ pair. The Coulomb interaction between $K^-$ and $p$
in this case makes some correction to the binding energy and width of the {$\Lambda(1405)$} resonance.

These two types of (quasi-)bound states in the $K^- p$ system are usually studied separately.  The $1s$ level shift
of kaonic hydrogen has been intensively studied experimentally and theoretically.  On the contrary, to the best
of our knowledge, no study has been devoted to changes in the  {mass} and width of the $\Lambda(1405)$ resonance
caused by an additional Coulomb interaction.  We aim to study both types of states simultaneously.

In the present paper, we study the shifts of the energy levels in the $K^- p$ system caused by both Coulomb
and strong antikaon-nucleon interactions.  We adopt simple strong $V_{\bar{K}N}$ potentials with real and
complex parameters fitted to the realistic antikaon-nucleon potential with coupled particle channels. We solved
the Lippmann-Schwinger equation in two different ways.

The first approach is an exact solution, which is possible for separable potentials. For a system described by
the Coulomb $V^{(C)}$ and strong $V^{(s)}$ potentials, the bound and quasi-bound states are the solutions of the 
homogeneous Lippmann-Schwinger equation 
\begin{equation}
\label{psiLS}
|\Psi\rangle = G^{(C)}(E) V^{(s)} |\Psi \rangle,
\end{equation}
where $G^{(C)}(E) = (E-H^{(0)}-V^{(C)})^{-1}$ is the Coulomb Green's operator and $H^{(0)}$ is the kinetic energy.
For a separable strong potential $V^{(s)}=|g\rangle\lambda\langle g|$ with Yamaguchi form-factors $|g\rangle$, 
the matrix elements of the Coulomb Green's operator $\langle g | G^{(C)} | g\rangle$ entering the equation for
the energy of the $i$th level can be calculated analytically, see e.g.~\cite{gGg_Coul1}. The wave function of the $i$th
level was evaluated from $\langle r | \Psi_i \rangle \sim \langle r |  G^{(C)}(E_i) | g \rangle$, using the coordinate
space representation of the Coulomb Green's function in terms of Whittaker functions.

Anticipating future work on the $K^- pp$ three-body system, we adopted a second approach, where we solved 
Eq.~ (\ref{psiLS})  in the Coulomb-Sturmian (CS) basis representation \cite{CS-method1}. The corresponding Coulomb
Green's matrix is given in an exact form in terms of $\mbox{}_2F_1$ hypergeometric functions. This method is equally
applicable for local and non-local potentials given in either momentum- or coordinate-space representation. 

Different antikaon-nucleon interaction models were used. We constructed one-channel simple versions of 
the coupled-channel  $V_{\bar{K}N - \pi \Sigma -\pi \Lambda}^{\rm 1,SIDD-A}$ potential from Ref.~\cite{FineTune_PRC} 
in particle representation $V_{K^- p}^{\rm 1A, real}$ and $V_{K^- p}^{\rm 1A, cmplx}$ 
\begin{equation}
  V_{K^- p}(k,k') = 
   \lambda_{K^- p} \, g(k) g({k'}), \quad g(k) = (k^2 + \beta^2)^{-1},
\label{VSep}
\end{equation}
which have real or complex strength constants $\lambda_{K^- p}$, respectively. The range parameter $\beta$
is the same for the real and complex versions of the simple potentials. They coincide with the $\beta_{1}$
value  of the coupled-channel $V_{\bar{K}N - \pi \Sigma -\pi \Lambda}^{\rm 1,SIDD-A}$ potential in the $I=0$
$\bar{K}N$ channel in Table II of Ref.~\cite{FineTune_PRC}. The $V_{K^- p}^{\rm 1A, real}$  and $V_{K^- p}^{\rm 1A, cmplx}$
potentials give the same  energy $E_1$ for the $\Lambda(1405)$ resonance as the coupled-channel one  (the real
version reproduces only its real part).  The parameters of our $K^- p$ potentials are presented in Table \ref{V_Kp.tab}.
%
\begin{table}[ht]
\begin{center}
\begin{tabular}{ccc}
\hline  \noalign{\smallskip}
  & $V_{K^- p}^{\rm 1A, real}$   & $V_{K^- p}^{\rm 1A, cmplx}$       \\[1mm]
\noalign{\smallskip} \hline
$\beta$ & $3.95$ & $3.95$ \\
 \noalign{\smallskip} \hline
$\lambda$ &   $-2.198$ & $-2.460 - i \, 0.537$  \\
 $E_1$ & $1429.2$ &  $1429.2 - i \, 33.5$ \\
 $a_{K^- p}$   & $-3.73$ & $-1.40 + i\,  0.86$ \\ [1mm]
 \noalign{\smallskip} \hline
\end{tabular}
\caption{
Range $\beta$ (fm$^{-1}$)  and strength $\lambda$ (fm$^{-2}$)  parameters
of the $K^- p$ potentials together with the ``strong" binding energies
$E_1$ (MeV) and $K^- p$ scattering lengths $a_{K^- p}$ (fm).  
\label{V_Kp.tab}
}
\end{center}
\end{table}

We calculated the lowest energy levels of the $K^- p$ system with both the strong and Coulomb interactions
and compared them with those  evaluated with the ``main" potentials only. The shifts of the levels are defined as
\begin{equation}
\label{Deltas_def}
\Delta E_0 = E_0^{s} - E_0^{s+C}, \quad \Delta E_i = E_i^{C} - E_i^{s+C}, \quad i = 1,2,\ldots,
\end{equation}
where $E_0$ is the energy of the ``strong" bound state ($\Lambda(1405)$), while $E_i$ defines the $i$th level
of the ``atomic" state, kaonic hydrogen. The top indices $s$, $C$, and $s+C$ denote the strong, Coulomb and
strong plus Coulomb potentials, respectively, used for the calculation of the energy. The energies $E$ are negative,
so $\Delta E < 0$ corresponds to the ``repulsive" shift of the level, when the additional interaction makes the level
less bound. The opposite case $\Delta E > 0$  is ``attractive."

The binding energies and level shifts of the ``strong" and lowest ``atomic" level energies of the $K^- p$ system
evaluated with strong $V_{K^- p}^{\rm 1A, real}$ and $V_{K^- p}^{\rm 1A, cmplx}$ potentials are shown in
Table~\ref{Kp_shifts.tab}. The results are quite unexpected.  Somewhat contradictorily,  the addition of the Coulomb
potential  to the strong $\bar{K}N$ potential shifts the ``strong" energy level much more that the addition of the strong
$\bar{K}N$ potential  to the Coulomb potential shifts the $1s$ and $2s$ atomic levels. The absolute value
of the $\Delta E_0$ caused by the addition of the weaker Coulomb interaction is larger than $\Delta E_1$ caused
by the addition of the stronger strong interaction by approximately three orders of magnitude!
%
\begin{table}[ht]
\begin{center}
\begin{tabular}{lcc}
\hline  \noalign{\smallskip}
  & $V_{K^- p}^{\rm 1A, real}$   & $V_{K^- p}^{\rm 1A, cmplx}$       \\[1mm]
\noalign{\smallskip} \hline
 $E_0^{s}$ (MeV) & $-5.37$ & $-5.37 - i \, 33.50$ \\
$E_0^{s+C}$ (MeV) & $-6.86$ &  $-7.68 -i \, 34.40$ \\
$\Delta E_0$ (MeV) & $1.49$ &  $2.31 + i \, 0.90$ \\
 \noalign{\smallskip} \hline
$E_{1}^{C}$ (keV) & $-8.64$ & $-8.64$ \\
 $E_1^{s+C}$ (keV) & $-7.62$ & $-8.13 - i \, 0.25$ \\
 $\Delta E_1$ (keV) & $-1.02$ & $-0.51 - i \, 0.25$ \\
\noalign{\smallskip} \hline
 $E_{2}^{C}$ (keV) & $-2.16$ & $-2.16$ \\
 $E_2^{s+C}$ (keV) & $-2.03 $ & $-2.10 -  i  \,   0.03$ \\
 $\Delta E_2$ (keV) & $-0.13$ & $-0.06 - i \, 0.03$ \\
 \noalign{\smallskip} \hline
\end{tabular}
\caption{
Energies of the $K^- p$ system $E_0$ (MeV), $E_1$ (keV), and $E_2$ (keV), and shifts of the levels $\Delta E$ from Eq.(\ref{Deltas_def}).
\label{Kp_shifts.tab}
}
\end{center}
\end{table}

Also confusing are the opposite signs of the ``strong" and ``atomic" level shifts in Table~\ref{Kp_shifts.tab}.
The energies of the $1s$ and $2s$ ``atomic" levels appear to move upwards in comparison to the pure Coulomb case, 
as if the additional strong $\bar{K}N$ interaction acts as a repulsive force. The long-standing kaonic hydrogen puzzle
concerned the paradoxical sign of the $1s$ level shift.  Since the attraction of the strong antikaon-nucleon interaction
is strong enough to lead to the existence of the quasi-bound state,  the $K^- p$ scattering length, extracted from
scattering experiments, is ``repulsive." Correspondingly, it was ``natural to expect that the strong-interaction level
shift for $K^- p$ atomic states \dots is also `repulsive' " \cite{BattyGal}. However, several earlier experiments reported
the attractive-type shift of the $1s$ level of the exotic atom.  Three recent experiments on kaonic hydrogen including
the most recent one \cite{SIDDHARTA}  improved the signal-to-background ratio and measured the ``repulsive" shift.
By this, it was assumed that the kaonic hydrogen puzzle was resolved.  But why then does the ``strong" $\Delta E_0$ 
shift caused by additional attractive Coulomb have the opposite sign?

Our answer is that $\Delta E_1$ and $\Delta E_2$ are not  the ``repulsive" ``atomic" shifts. In fact, they are not the shifts,
but differences between neighbor levels of the $K^- p$ system. What we see is level rearrangement.
\begin{figure}
\includegraphics[width=0.85\columnwidth]{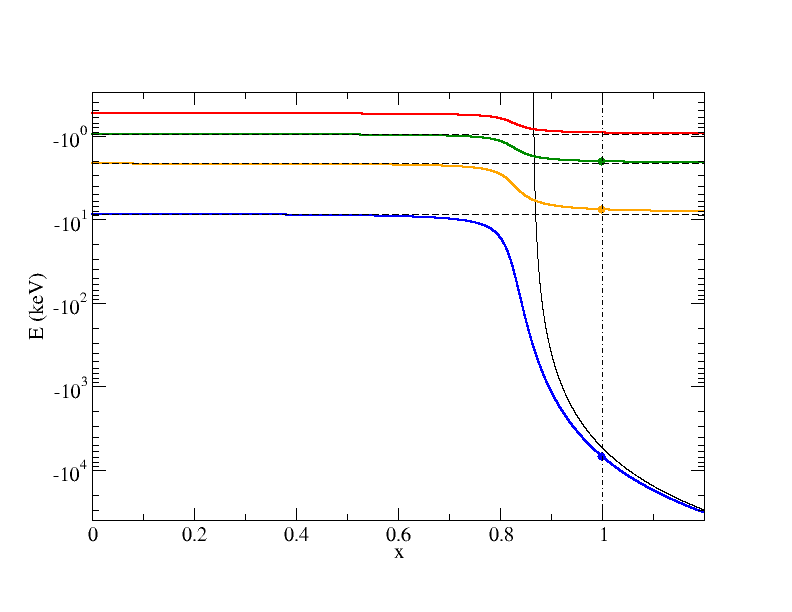}
\caption{Level rearrangement in the $K^- p$ system calculated with $V_{K^- p}^{\rm 1A, real}$ potential with
gradually changing real strength constant $\lambda_x = x\lambda$.
\label{Zeld_real.fig}
}
\end{figure}

It was found in Ref.~\cite{Zeldovich_orig} that a strong interaction added to a system formed by a pure Coulomb
potential leads to level rearrangement for some critical value of the interaction, corresponding to the value at which
the pure strong bound state appears. As the attraction of the strong potential increases,  after the critical value
the lowest $1s$ state of the atom follows the trajectory of the pure strong state which is much stronger than
the  $1s$ level Coulomb energy. At the same point, the $2s$ level of the atom drops down and occupies almost
the energy which was the $1s$ level of the original pure Coulomb system. The third atomic level moves toward
the initial $2s$ and so on. It is exactly what is seen at Fig.~\ref{Zeld_real.fig}.

In the figure, the trajectories of the levels in the $K^- p$ system (solid lines)  were evaluated with the gradually changing
strength constant of the  $V_{K^- p}^{\rm 1A, real}$ strong potential  $\lambda_x = x\lambda$, with $x\in [0,1.2]$.
At $x = 0$, these trajectories coincide with the energy levels of kaonic hydrogen calculated with the Coulomb potential
only (dashed horizontal lines). The solid black line denotes the trajectory of the pure ``strong" energy level, which does
not exist at smaller values of the strength constant $\lambda_x$. The value $x=1$ corresponding to $\lambda_x = \lambda$
from Table \ref{V_Kp.tab} is denoted by the vertical dash-dotted line.  The energies calculated with both interactions
at $\lambda_x = \lambda$ are denoted by circles.

It is seen from Fig.~\ref{Zeld_real.fig} that all solid lines are gradually moving downwards when the absolute value
of the strength constant of the strong interaction $\lambda_x$ grows. The physical value of the lowest level calculated
with Coulomb and strong interactions (blue circle) is the shifted level, which at $x=0$, i.e., without strong
interaction, coincides with the  pure Couomb $1s$ level. On the other hand, the same energy  is the pure ``strong" energy
level (black line) also shifted downwards. Therefore, both the ``atomic" and ``strong" shifts are ``attractive".

The same is true for the first excited state of the $K^- p$ system (orange circle). It is seen from Fig.~\ref{Zeld_real.fig}
that this energy is the shifted $2s$ level calculated with Coulomb interaction only. This shift is also ``attractive". 
In contrast to the ground state, there is only one type of shift for all excited states, namely, the downward shiftcaused by the additional strong interaction.

\begin{figure}
\includegraphics[width=0.90\columnwidth]{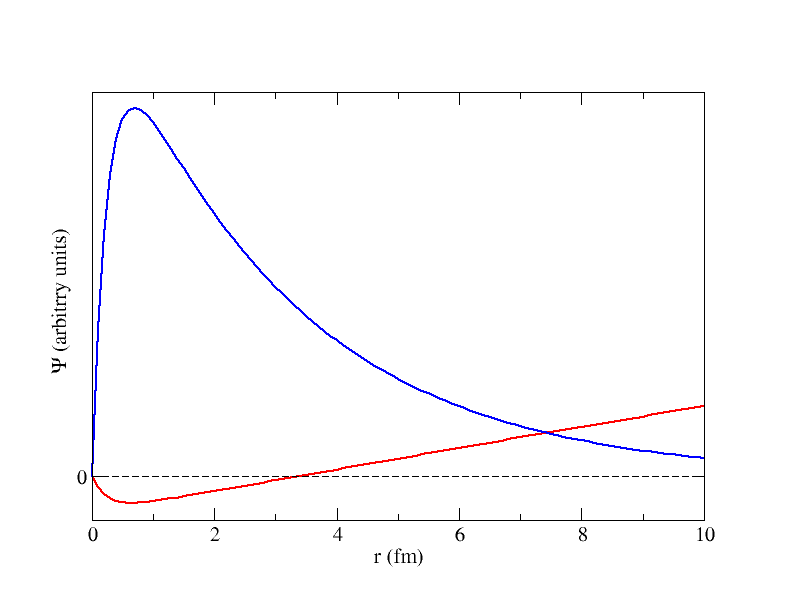}
\caption{Wave functions of the ground $1s$ (blue line) and first excited $2s$ (multiplied by $10$, red line) states of the
$K^- p$ system, $V_{K^- p}^{\rm 1A, real}$ potential.
\label{wf_real.fig}
}
\end{figure}

An additional proof that the $E_{2s}^{s+C}$ energy is the energy of the first exited level of the system and not
of the ground state is its wave function, shown in Fig.~\ref{wf_real.fig}. Namely, the $2s$ wave function 
(red line, multiplied by $10$) has a node close to the origin, while the wave function of the ground $1s$ state 
(blue line) does not have any nodes.

In practice, the $1s$ level shift of kaonic hydrogen  is calculated as the difference between the pure Coulomb $1s$
energy level (the lowest dashed line) and the value of the closest to it shifted energy, denoted as the orange circle in Fig.~\ref{Zeld_real.fig}. However, we see that this energy difference is not the level shift, but the difference between
the pure Coulomb $1s$ and the first exited $2s$ energy levels of the $K^-p$ system calculated with both interactions.
The correctly evaluated shift of the $1s$ level is the difference between the pure Coulomb $1s$ level and the lowest
value of the energy of the $K^- p$ system (the blue circle in Fig.~\ref{Zeld_real.fig}). The same is true for the  excited states
of the system. The correct definition of the $1s$ and $2s$ level shifts is
\begin{eqnarray}
\nonumber
\label{CorrDeltas}
\Delta E_{1s}^{s} &=& E_{1s}^{C} - E_{1s}^{s+C}, \qquad \Delta E_{1s}^{C} = E_{1s}^{s} - E_{1s}^{s+C},  \\
\Delta E_{2s}^{s} &=& E_{2s}^{C} - E_{2s}^{s+C},
\end{eqnarray}
where $\Delta E^{s}$ defines the shift caused by the strong interaction for different levels, and $\Delta E_{1s}^{C}$
is the shift of initially pure strong level caused by the Coulomb potential. The correct shifts of the ground $1s$ and
first exited $2s$ states are presented in Table~\ref{CorrShifts.tab}.
%
\begin{table}[ht]
\begin{center}
\begin{tabular}{lcc}
\hline  \noalign{\smallskip}
  & $V_{K^- p}^{\rm 1A, real}$   & $V_{K^- p}^{\rm 1A, cmplx}$       \\[1mm]
\noalign{\smallskip} \hline
 $E_0^{s}$ (MeV) & $-5.37$ & $-5.37 - i \, 33.50$ \\
$E_{1s}^{s+C}$ (MeV) & $-6.86$ &  $-7.68 -i \, 34.40$ \\
$\Delta E_{1s}^{C}$ (MeV) & $1.49$ &  $2.31 + i \, 0.90$ \\
 \noalign{\smallskip} \hline
$E_{1s}^{C}$ (MeV) & $-0.00864$ & $-0.00864$ \\
$E_{1s}^{s+C}$ (MeV) & $-6.86$ &  $-7.68 -i \, 34.40$ \\
$\Delta E_{1s}^{s}$ (MeV) & $6.85$ &  $7.67 + i \, 34.40$ \\
\noalign{\smallskip} \hline
 $E_{2s}^{C}$ (keV) & $-2.16$ & $-2.16$ \\
 $E_{2s}^{s+C}$ (keV) & $-7.62$ & $-8.13 - i \, 0.25$ \\
 $\Delta E_{2s}^s$ (keV) & $5.46$ & $5.97 + i \, 0.25$ \\
 \noalign{\smallskip} \hline
\end{tabular}
\caption{The shift of the ``strong" level $\Delta E_{1s}^C$ and
correctly evaluated level shifts of the $1s$ and $2s$ ``atomic" levels of the $K^- p$ system
$\Delta E_{1s}^{s}$, $\Delta E_{2s}^{s}$.
\label{CorrShifts.tab} 
}
\end{center}
\end{table}

It is seen from Table ~\ref{CorrShifts.tab} that the $1s$ level shift of kaonic hydrogen $\Delta E_{1s}^{s}$ is 
positive, so it is not ``repulsive". The shifts of all levels are ``attractive" as is expected from the addition of
an attractive interaction. The correctly evaluated shift of the $1s$ level of kaonic hydrogen $\Delta E_{1s}^{s}$
caused by the additional strong interaction is large, and it is much larger than the shift of the $\Lambda(1405)$
resonance $\Delta E_{1s}^{C}$ caused by the additional Coulomb potential.  It is what we expected, keeping in mind
the different strengths of the interactions:  the weaker Coulomb potential leads to a smaller shift.

The above discussion concerns a strong $V_{K^- p}^{\rm 1A, real}$ potential with real strength parameter $\lambda$. 
However, the strong $K^- p$ interaction is absorptive, and it was argued in  Refs.~\cite{Zeldovich_absorpt,Zeldovich_gal}
that  sufficiently strong absorption in the strong interaction can destroy a  level rearragement effect. Therefore, we also
evaluate the energy levels and wave functions in the $K^- p$ system using the complex antikaon-nucleon potential.
%
\begin{figure}
\includegraphics[width=0.85\columnwidth]{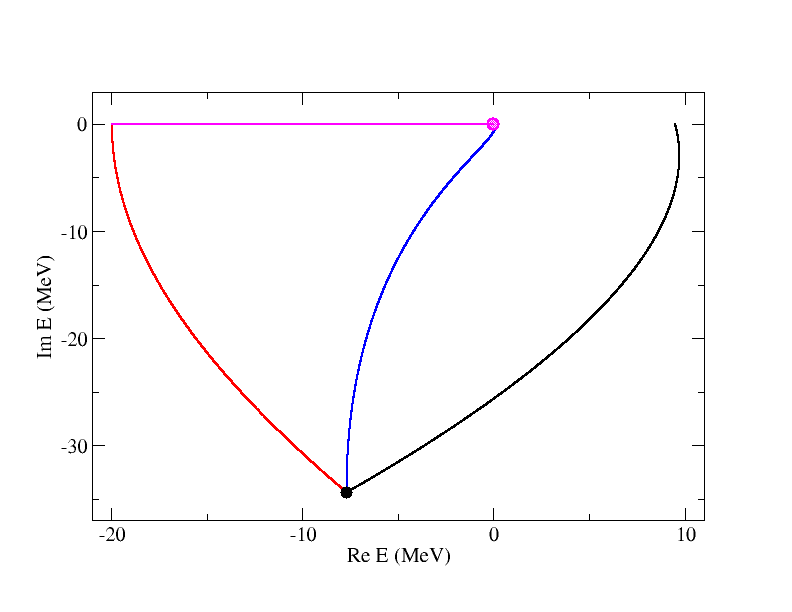}
\caption{
Level rearrangement in the $K^- p$ system calculated using $V_{K^- p}^{\rm 1A, cmplx}$ potential with 
gradually changing complex strength constant along the $\lambda_x^{(1)}$ (black line), $\lambda_x^{(2)}$
(red and magenta lines), and  $\lambda_x^{(3)}$ (blue line) trajectories.
\label{Zeld_cmplx.fig}
}
\end{figure}

The wave function with the energy $E_{2s}^{s+C}$ calculated with the strong $V_{K^- p}^{\rm 1A, cmplx}$ 
potential has an additional node, similar to the case of the real $K^- p$ potential. It means that the state has
become the first excited state of the system, and  the level rearrangement effect takes place.

It turned out that the visibility of the level rearrangement effect for a strong potentials with absorption depends
on the trajectory in the complex energy plane, as is seen in Fig.~\ref{Zeld_cmplx.fig}. When the most natural way
of changing the strength constant $\lambda_x^{(1)} = x \, \Re \lambda + i \, x \, \Im \lambda$ (black lines), $x = 1 \dots 0$
is used, the lowest energy level $E_{1s}^{s+C}$ calculated with the strong $V_{K^- p}^{\rm 1A, cmplx}$ and Coulomb
potentials (black circle in the figure) moves to an unphysical Riemann sheet. It means that the level rearrangement
is not seen in this case.

The second trajectory $\lambda_x^{(2)} $ consists of two parts. First, the magnitude of $\Im \lambda_x$ gradually
decreases to zero with fixed real part (red line), and then the magnitude of $\Re \lambda_x$ decreases to zero with
$\Im \lambda_x = 0$ (magenta line).  The poles on the third trajectory were calculated with 
$\lambda_x^{(3)} = x \, \Re \lambda + i \, x^7 \, \Im \lambda$  (blue line). It is seen in Fig.~\ref{Zeld_cmplx.fig} that
trajectories 2 and 3 lead the $E_{1s}^{s+C}$ pole to the pure Coulomb $1s$ energy $E_{1s}^{C}$ (magenta circle).
The behavior of the real parts of the energy in these last cases is the same as the one calculated with the real strong
interaction model $V_{K^- p}^{\rm 1A, real}$.  It means that now we see the level rearrangement in the system with absorption. 

Therefore, all the conclusions we have drawn for the real strong antikaon-nucleon potential are valid for the complex
model as well. The correct values of the level shifts can be found in Table~\ref{CorrShifts.tab}. The shifts of all levels
are positive, i.e., they are ``attractive", and the shift of the $1s$ level caused by the strong interaction is larger than
that caused by the additional Coulomb potential.

It is also interesting that $\Delta E_{1s}^{C}$  being the Coulomb correction to the $\Lambda(1405)$ resonance
mass is not so small compared to the binding energy of the (quasi-)bound state: it shifts the energy by about $28 \%$ for
the real and $43 \%$ for the complex antikaon-nucleon potentials.

From an experimental point of view, it seemed that $2p \to 1s$ gamma-ray transition was measured. However, it is rather
$2p \to 2s$ transition, which is absent in the case of pure Coulomb interaction since these two states are degenerate. 
However, in the presence of the strong antikaon-nucleon interaction, as we see, the $1s$ and $2s$ levels move downwards,
while the $2p$ level stays almost unchanged (for a local potential) or exactly the same (for an $s$-wave separable one).
Thus, instead of being the $1s$ level shift defined by
$\Delta_{1s} = E^{exp}_{2p \to 1s} - E^{C}_{2p \to 1s} \approx - E^{exp}_{1s} + E^{C}_{1s}
$
the value presented as the experimental result is rather the difference between the neighboring levels
$
\Delta_{2s - 1s} = E^{exp}_{2p \to 2s} - E^{C}_{2p \to 1s} \approx - E^{exp}_{2s} + E^{C}_{1s} 
$

Summarizing, we studied different types of energy levels in the $K^- p$ system and their shifts. We demonstrated
that the level rearrangement effect takes place in the system when the simple real and complex $\bar{K}N$ strong
potentials are used together with the Coulomb interaction.

We showed that the $1s$ level shift of kaonic hydrogen is not ``repulsive," but all the shifts caused by the additional
strong or Coulomb interactions are ``attractive". The correct way of calculating the shifts is defined by 
Eq. (\ref{CorrDeltas}).  The measured and theoretically calculated values are really the differences between the $2s$
level of the whole $K^- p$ system and the pure Coulomb $1s$ energy level.

The correctly calculated shift of the $1s$ level of kaonic hydrogen $7-8$ MeV is huge  compared to the pure Coulomb
energy.  The shift of the $\Lambda(1405)$ caused by the addition of the Coulomb potential is smaller, about $1-2$ MeV.
However, it is not so small compared to the binding energy of this quasi-bound state, being the ground state of
the $K^- p$ system. The shift of the $2s$ level of kaonic hydrogen caused by the strong interaction is much smaller,
$\sim 5-6$ keV. 

\bibliography{kp}
\bibliographystyle{apsrev4-2}

\end{document}